\documentstyle [12pt] {article}
\begin{document}

\pagestyle{empty}
\rightline{\vbox{
\halign{&#\hfil\cr
&FERMILAB-PUB-94/074-T\cr
&NUHEP-TH-94-6\cr
&March 1994\cr}}}
\bigskip
\bigskip
\bigskip
{\Large\bf
	\centerline{$J/ \psi$ Production from}
        \centerline{Electromagnetic Fragmentation}
	\centerline{in $Z^{0}$ decay}}
\bigskip
\normalsize
\centerline{Sean Fleming}
\centerline{\sl Department of Physics and Astronomy}
\centerline{\sl Northwestern University,Evanston, IL 60208}
\centerline{and}
\centerline{\sl Fermi National Accelerator Laboratory}
\centerline{\sl P.O. Box 500, Batavia, IL 60510}
\bigskip

\begin{abstract}
The rate for $\; Z^{0}\rightarrow J/ \psi + \ell^{+}\ell^{-} \;$ is suprisingly
large with about one event for every million $Z^{0}$ decays. The reason for
this is that there is a fragmentation contribution that is not
suppressed by a factor of $M^{2}_{\psi}/M^{2}_{Z}$. In the fragmentation limit
$\; M_{Z}\rightarrow\infty$ with $E_{\psi}/M_{Z}$ fixed, the differential
decay rate for $\;Z^{0}\rightarrow J/ \psi + \ell^{+}\ell^{-} \;$ factors into
electromagnetic decay rates and universal fragmentation functions. The
fragmentation functions for lepton fragmentation and photon fragmentation into
$J/\psi$ are calculated to lowest order in $\alpha$. The fragmentation
approximation to the rate is shown to match the full calculation for $E_{\psi}$
greater than about $3 M_{\psi}$.

\end{abstract}
\vfill\eject\pagestyle{plain}\setcounter{page}{1}
{\bf\centerline{Introduction}}

Fragmentation is the decay of a high transverse momentum parton into a
collinear
hadron. The differential cross section for the inclusive production of such a
hadron in $e^{+} e^{-}$ annihilation factors into differential cross
sections $d \hat{\sigma}$ for the production of large transverse momentum
partons and fragmentation functions $D(z)$ \cite{cs}. The fragmentation
function gives the probability for the splitting of a parton into the hadron
with momentum fraction $z$. These functions are independent of the subprocess
that creates the fragmenting particle, and can be evolved to any scale via the
Altarelli-Parisi evolution equations.

It has recently been shown that it is possible to calculate the fragmentation
functions for heavy quarkonium states using perturbative quantum chromodynamics
(QCD) \cite{by}. Fragmentation functions for several of these states have been
calculated explicitly \cite{by,bcy:cfrag,bcy:bfrag,flsw,cc}. In particular the
paper of Braaten, Chueng and Yuan on charm quark fragmentation \cite{bcy:cfrag}
is most relevant to the work presented here. Their analysis focuses on decays
of the $Z^{0}$ into hadrons showing that charmonium production is dominated by
charm quark fragmentation. In an analagous manner it is shown here that
decays of the $Z^{0}$ into charmonium plus electromagnetic particles (leptons
and photons) are dominated by electromagnetic fragmentation.

The process $Z^{0}\rightarrow \psi + \ell^{+} \ell^{-}$, which is of
order $\alpha^2$, where $\alpha$ is the electromagnetic coupling constant, has
a branching ratio of $7.5 \times 10^{-7}$. This is an order of magnitude
larger than the order-$\alpha$ process $Z^{0} \rightarrow\psi \gamma$
\cite{gkpr}, which has a branching ratio of $5.2\times 10^{-8}$. That makes
$Z^{0}\rightarrow \psi + \ell^{+} \ell^{-}$ the dominant $\psi$ production
mechanism in electromagnetic $Z^{0}$ decays.  The unexpectedly large rate can
be explained by a fragmentation contribution which is not suppressed by a
factor of $M^{2}_{\psi}/M_{Z}^{2}$ \cite{by}.  In this paper, it is shown that
in the fragmentation limit $\;M_{Z}\rightarrow\infty$ with
$E_{\psi}/M_{Z}$ fixed, the rate for $Z^{0}\rightarrow \psi +
\ell^{+}\ell^{-}$ factors into subprocess rates for electromagnetic
decays and individual fragmentation functions . At leading order in
$\alpha$ there is a contribution from the fragmentation function
$D_{\ell \rightarrow \psi}$ for a lepton to split into a $\psi$, and a
contribution from the fragmentation function $D_{\gamma \rightarrow
\psi}$ for a photon to split into a $\psi$. The fragmentation
functions are calculated in a manner that is independent of the hard
process that produces the fragmenting parton, and the fragmentation
calculation is shown to match the full calculation for $E_{\psi}$ greater than
about $3 M_{\psi}$.

\bigskip
{\bf \centerline{The decay rate for $Z^{0} \rightarrow \psi +\; \ell^{+}
\ell^{-}$}}

The decay rate for $Z^{0}\rightarrow\psi \; + \; \ell^{+}\ell^{-}$ was
calculated in a model-independent way in Ref. \cite{flem,br} using the Feynman
diagrams in figure 1a. The diagrams in figure 1b also contribute to this
process at the same order in the electromagnetic coupling $\alpha$.
Fortunately they can be neglected.

In order to understand why only the diagrams of figure 1a need to be considered
it is necessary to understand why the rate for $Z^{0} \to \psi \gamma$ is
smaller than the rate for $Z^{0} \to \psi + \ell^{+} \ell^{-}$. Naively one
would expect $\Gamma(Z^{0} \to \psi \gamma)$ to be larger than $\Gamma(Z^{0}
\to \psi + \ell^{+} \ell^{-})$ since the latter rate is suppressed by a power
of $\alpha$ compared to the former rate. However upon closer examination this
expectation turns out to be untrue. To see why compare the diagrams that
contribute to the decay rates $\Gamma(Z^{0}\to \psi + \ell^{+} \ell^{-})$
figure 1 and $\Gamma(Z^{0} \to \psi \gamma)$ figure 2. It is important to note
that the lepton propagator in the diagrams of figure 2 is always of order
$1/M^{2}_{Z}$. In contrast there is a substantial region of phase space where
both the photon propagator and the lepton propagator in the diagrams of figure
1a are of order $1/M^{2}_{\psi}$. Thus these diagrams will be enhanced by a
factor of $M^{2}_{Z}/M^{2}_{\psi}$, compared to the diagrams of figure 2. The
factor of $M^{2}_{Z}/M^{2}_{\psi}$ is large enough to overwhelm the extra
power of $\alpha$ making the diagrams of figure 1a more important than the
diagrams of figure 2. The lepton propagator of the diagrams in figure 1b is
always of order $1/M^{2}_{Z}$, so these diagrams are suppressed by a factor of
$\alpha$ compared to the diagrams in figure 2, and may be neglected.

The remaining diagrams that contribute to
$\Gamma(Z^{0} \to \psi + \ell^{+} \ell^{-})$ at the same order in $\alpha$ are
obtained from the diagrams of figure 1 by replacing the photon propagator with
a $Z$-boson. Then the boson propagator in the diagrams of figure 1a is always
of order $1/M^{2}_{Z}$. Similarly the lepton propagator in the diagrams of
figure 1b is also always of order $1/M^{2}_{Z}$. Thus these diagrams are
suppressed by a factor of $\alpha$ compared to the diagrams of figure 2 and can
be neglected.

Keeping only the diagrams in figure 1a, and neglecting the lepton mass, the
result of the calculation of
$\Gamma (Z^{0}\rightarrow\psi + \ell^{+}\ell^{-})$ is
\begin{eqnarray}
\Gamma & = &
4 \alpha^{2} g^{2}_{\psi}\;  \Gamma (Z^{0}\rightarrow\ell^{+}\ell^{-})
\int^{1+ \lambda}_{2 \sqrt{\lambda}}dy
\int^{\sqrt{y^{2}-4\lambda}}_{-\sqrt{y^{2}-4\lambda}} dw \; \;
g(y,w,\lambda)
\nonumber \\
 g(y,w,\lambda) & = & \frac{1}{2}\left(\frac{(y-2)^{2}+w^{2}}{y^{2}-w^{2}}
\right) - 2 \lambda \left(\frac{2w^{2}-(y^{2}-w^{2})(1-y)}{(y^{2}-w^{2})^{2}}
\right) + 2\lambda^{2} \! \left(\frac{1}{y^{2}-w^{2}}\right)
\label{eq:rate}
\end{eqnarray}
where $\lambda=M_{\psi}^{2}/M_{Z}^{2}$, $y=2 P \cdot Z/M^{2}_{Z}$ and
$w=2(p_{+}-p_{-}) \cdot Z/M^{2}_{Z}$. Here $P$, $\; p_{-}$, $\; p_{+}$, and
$Z$ are the 4-momenta of the $\psi$, $\; \ell^{-}$, $\; \ell^{+}$, and $Z^{0}$.
The parameter $g_{\psi}$ can be determined from the electronic width
$\Gamma_{e^{+}e^{-}}$ of the $\psi$ to be
\begin{equation}
g^{2}_{\psi}=\frac{3}{4 \pi}
\frac{\Gamma_{e^{+}e^{-}}}{\alpha^{2} M_{\psi}}.
\label{g}
\end{equation}
Using $\Gamma_{e^{+}e^{-}} = 5.4$ keV the photon-to-$\psi$ coupling is
$g^{2}_{\psi}=0.008$. Integrating the function $g(y,w,\lambda)$ over $w$
yields the full differential decay rate
\begin{eqnarray}
\lefteqn{\frac{d \Gamma}{d E_{\psi}}=}
\nonumber \\
& & 8 \alpha^{2} g^{2}_{\psi} \;
\frac{\Gamma (Z^{0}\rightarrow\ell^{+}\ell^{-})}{M_Z} \left[ \left(
\frac{(y-1)^2+1}{y} + \; \lambda \; \frac{3-2y}{y} \; + \; \lambda^{2} \;
\frac{2}{y} \right)\log{\frac{y+y_{L}}{y-y_{L}}} \; - \; 2 y_{L} \right],
\label{eq:gfun}
\end{eqnarray}
where $y_{L} = \sqrt{y^{2}-4\lambda}$. In the center of mass frame $y=2
E_{\psi}/M_{Z}$.
Later on these results will be compared to the fragmentation calculation in
the fragmentation limit $\; M_{Z}\rightarrow\infty$ with $E_{\psi}/M_{Z}$
fixed. In this limit Eq. (\ref{eq:gfun}) reduces to
\begin{equation}
\frac{d \Gamma}{d E_{\psi}} = 8 \alpha^{2} g^{2}_{\psi} \;
\frac{\Gamma (Z^{0}\rightarrow\ell^{+}\ell^{-})}{M_Z}
\left[ \frac{(y-1)^2+1}{y} \log{\frac{y^2}{\lambda}} \; - \; 2 y \right].
\label{xxx}
\end{equation}

\bigskip
{\bf \centerline{The Fragmentation Contribution to $Z^{0}$ Decay}}

In reference \cite{bcy:cfrag} the general form of the fragmentation
contribution for the production of a $\psi$ of energy $E_{\psi}$ in $Z^{0}$
decays is given as
\begin{equation}
d \Gamma (Z^{0} \rightarrow \psi (E_{\psi})+X) =
 \sum_{i} \int dz \; d\widehat{\Gamma}
(Z^{0}\rightarrow i(E_{\psi}/z)+X,\mu^{2})\; D_{i\rightarrow \psi}(z,\mu^{2}),
\label{eq:frag}
\end{equation}
where the sum is over partons and $z$ is the longitudinal momentum fraction
of the $\psi$ relative to the fragmenting parton. All of the dependence on the
$\psi$ energy $E_{\psi}$ has been factored into the subprocess decay rate
$\widehat{\Gamma}$ and all of the dependence on the $\psi$ mass $M_{\psi}$
has been factored into
the fragmentation function $D_{i \rightarrow \psi}(z,\mu^{2})$. A factorization
scale $\mu$ has to be introduced in order to maintain this factored form in all
orders of perturbation theory. This general form was developed in the context
of QCD, but it applies equally to QED, where the only partons are leptons and
photons. This simplifies the general electromagnetic fragmentation contribution
to
\begin{eqnarray}
\lefteqn{\frac{d \Gamma}{dE_{\psi}} (Z^{0} \rightarrow \psi (E_{\psi})+X)=}
\nonumber \\
& & 2\int^{1}_{0} dz \int dE_{\ell} \; \frac{d \widehat{\Gamma}}{dE_{\ell}}
(Z^{0}\rightarrow\ell^{-} (E_{\ell})+X,\mu^{2}) \;
D_{\ell\rightarrow\psi}(z, \mu^{2})\; \delta(E_{\psi}-zE_{\ell})
\nonumber \\
& & + \int^{1}_{0} dz \int dE_{\gamma} \; \frac{d
\widehat{\Gamma}}{dE_{\gamma}}
(Z^{0}\rightarrow \gamma(E_{\gamma})+X, \mu^{2})\;
D_{\gamma\rightarrow\psi}(z,\mu^{2}) \; \delta(E_{\psi}-zE_{\gamma}),
\label{qedfrag}
\end{eqnarray}
where $X$ are electromagnetic final states, and $E_{\ell}$ and
$E_{\gamma}$ are the lepton and photon energies. The factor of 2
accounts for the fragmentation contribution from both the $\ell^{-}$
and the $\ell^{+}$. Large logarithms of $E_{\psi}/\mu$ in the
subprocess decay rate $\widehat{\Gamma}$ can be avoided by choosing
$\mu$ on the order of $E_{\psi}$. The large logarithms of order
$E_{\psi}/M_{\psi}$ which then appear in the fragmentation functions
can be summed up by solving the Altarelli-Parisi evolution equation.
In the electromagnetic case the evolution of the fragmentation functions
is of order $\alpha$ and may be neglected.

It is easy to count the order of $\alpha$ for the leading order
fragmentation contributions to Eq. (\ref{qedfrag}). The subprocess
rate for $Z^{0} \rightarrow \ell^{-} + X$ is of order 1, while the
subprocess rate for $Z^{0} \rightarrow \gamma + X$ is of order
$\alpha$. The fragmentation function for a lepton to split into a
$\psi$ will be shown to be of order $\alpha^{2}$, while the
fragmentation function for a photon to split into a $\psi$ will be
shown to be of order $\alpha$. Therefore both fragmentation processes
will contribute to Eq. (\ref{qedfrag}) at leading order in $\alpha$.

At lowest order in $\alpha$ it is possible to simplify things. The energy
distribution for the subprocess $Z^{0} \to \ell^{-}(E_{\ell}) + X$ at lowest
order is
\begin{equation}
\frac{d \widehat{\Gamma}}{d E_{\ell}} = \widehat{\Gamma}
(Z^{0} \to \ell^{+} \ell^{-}) \; \delta(E_{\ell} - \frac{M_{Z}}{2}).
\label{sim}
\end{equation}
Furthermore the photon fragmentation function at lowest order can be written as
\begin{equation}
D_{\gamma \to \psi}(z,  \mu^2) = P_{\gamma \to \psi} \; \delta(z-1)
\label{fo}
\end{equation}
where the function $P_{\gamma \to \psi}$ is the probability for a photon to
split into a $\psi$, and $z$ is the longitudinal momentum fraction of the
$\psi$
relative to the $\gamma$. These simplifications reduce the fragmentation
contribution to the energy distribution Eq. (\ref{qedfrag}) at leading order
in $\alpha$ to:
\begin{eqnarray}
\lefteqn{ \frac{d\Gamma}{dE_{\psi}}(Z^{0}\rightarrow \psi(E_{\psi}) + \;
\ell^{+}\ell^{-})= }
\nonumber \\
& & \frac{4}{M_{Z}} \; \widehat{\Gamma}(Z^{0}\rightarrow\ell^{+}\ell^{-}) \;
D_{\ell\rightarrow\psi}(\frac{2E_{\psi}}{M_{Z}}, \mu^{2}) \; + \;
\frac{d\widehat{\Gamma}}{dE_{\psi}}(Z^{0}\rightarrow \gamma (E_{\psi})+
\ell^{+}\ell^{-}, \mu^{2})
 \; P_{\gamma\rightarrow\psi}   .
\label{eq:lofrag}
\end{eqnarray}
The physical interpretation of the first term on the right hand side of
Eq. (\ref{eq:lofrag}) is that the $Z^{0}$ decays into two leptons, with
energies $M_{Z}/2$ on a distance scale of order $1/M_{Z}$. Subsequently one of
the leptons decays into a collinear lepton and $\psi$ on a distance scale of
order $1/M_{\psi}$. The physical interpretation of the second term is that the
$Z^{0}$ decays into two leptons and a photon with energy $E_{\psi}$ on a
distance scale of order $1/M_{Z}$, and the photon fragments into a $\psi$ on a
distance scale of order $1/M_{\psi}$. Note that at this order the only
dependence on the factorization scale is in $D_{\ell \rightarrow \psi}$ and in
the subprocess rate for $Z^{0} \rightarrow \gamma +\ell^{+} \ell^{-}$.

Given the general form of the fragmentation contribution at lowest order in
$\alpha$ in Eq. (\ref{eq:lofrag}), it is only necessary to calculate the
fragmentation function $D_{\ell \rightarrow \psi}(z, \mu^{2})$, the
fragmentation probability $P_{\gamma\rightarrow\psi}$, and the subprocess rate
for the $Z^{0}\rightarrow \gamma +\ell^{+}\ell^{-}$.

\bigskip
{\bf \centerline{Photon Fragmentation}}

The fragmentation function $D_{\gamma\rightarrow\psi}(z,\mu^{2})$ for a
photon to split into a $\psi$ can be calculated in a manner that is independent
of the process that produces the fragmenting photon. The Feynman diagram in
figure 3a represent such a process at lowest order in $\alpha$. An unknown
vertex, represented by the circle, radiates a photon which fragments into a
$\psi$. The fragmentation probability $ P_{\gamma\rightarrow\psi}$ can be
isolated by dividing the cross section $\sigma_{1}$, for the production of a
$\psi$ with energy $E_{\psi}$, by the cross section $\sigma_{0}$, for the
production of a real photon of energy $E_{\gamma}=E_{\psi}$, in the limit
$E_{\psi} \gg M_{\psi}$ where fragmentation dominates.

The general form of the photon production cross section $\sigma_{0}$ is
\begin{equation}
\sigma_{0}= \frac{1}{Flux} \int [dk] [dp_{out}] \; (2 \pi)^{4}
\delta^{4}(p_{in}-k-p_{out})\; \sum |A_{0}|^{2}
\label{s0}
\end{equation}
where $p_{in}$ is the sum of incoming 4-momenta, $k$ is the photon
4-momentum, and $p_{out}$ is the sum of the remaining outgoing
4-momenta. Here $[dk]=d^{3}k/(16 \pi^{3}k_{0})$ is the
Lorentz-invariant phase space for the photon and $[dp_{out}]$ is the
Lorentz invariant phase space for the remaining outgoing particles,
$Flux$ denotes the incoming particle flux for which no explicit
expression is needed since it will cancel the same factor when
$\sigma_{1}$ is divided by $\sigma_{0}$. The amplitude $A_{0}$ for the
process can be calculated from the Feynman diagram in figure 3b
\begin{equation}
A_{0}= \Gamma^{\mu} \; \epsilon_{\mu},
\label{amp0}
\end{equation}
where $\Gamma^{\mu}$ is a vertex factor for the production of the
photon, for which the explicit form is not needed. Squaring and
summing over final spins gives
\begin{equation}
\sum |A_{0}|^{2}=-\Gamma^{\mu}\Gamma^{*}_{\mu}.
\label{ampsq}
\end{equation}

The general form of the $\gamma \rightarrow \psi$ cross section $\sigma_{1}$ is
\begin{equation}
\sigma_{1}= \frac{1}{Flux} \int [dP] [dp_{out}]\; (2 \pi)^{4}
\delta^{4}(p_{in}-P-p_{out}) \; \sum{|A_{1}|}^{2}
\label{s1}
\end{equation}
where $P$ is the $\psi$ 4-momentum. $Flux$, $p_{in}$, and $p_{out}$ are the
same as described after Eq. (\ref{s0}). The amplitude $A_{1}$ can be
calculated from the Feynman diagram in figure 3a
\begin{equation}
A_{1}=
\frac{g_{\psi} e}{4}
\; \Gamma^{\mu} \; (-g_{\mu \nu}) \; {\rm tr}[\not{ \! \epsilon} \gamma^{\nu}],
\label{amp1}
\end{equation}
where $\epsilon$ is the $\psi$ polarization vector, and $\Gamma^{\mu}$ is the
vertex factor for the production of a virtual photon with invariant mass
$M_{\psi}$. Squaring the amplitude and summing over final spins gives
\begin{equation}
\sum |A_{1}|^{2} = -4 \pi g^{2}_{\psi} \alpha \; \Gamma^{\mu}\Gamma^{*}_{\mu}
\label{ampsq1}
\end{equation}
The vertex factor $\Gamma^{\mu}$ in Eq. (\ref{amp1}) only differs from the
vertex factor in Eq. (\ref{amp0}) in one respect, it is evaluated at the point
$k^{2}=M^{2}_{\psi}$ instead of $k^{2}=0$. Aside from the vertex factor the
$\psi$ mass enters the cross section $\sigma_{1}$ in the phase space integral.
The contribution to photon fragmentation will come from a region of phase space
where $E_{\psi} \gg M_{\psi}$ so that the $\psi$ mass can be neglected. Then
the cross section $\sigma_{1}$ can be written as
\begin{equation}
\sigma_{1}= 4 \pi  g^{2}_{\psi} \alpha \; \sigma_{0}.
\label{fragcs}
\end{equation}
The fragmentation probability $P_{\gamma\rightarrow\psi}$ can be read off from
Eq.(\ref{fragcs}):
\begin{equation}
P_{\gamma\rightarrow\psi} = 4 \pi g^{2}_{\psi} \alpha.
\label{photprob}
\end{equation}
Evaluating this numerically gives a fragmentation probability of
$P_{\gamma\rightarrow\psi} = 7 \times 10^{-4}$.

\bigskip
{\bf\centerline{Photon Subprocess}}

The energy distribution of the subprocess $Z^{0}\rightarrow \gamma
+\ell^{+}\ell^{-}$ is calculated next.
At this point it is necessary to decide what part of the
$Z^{0} \rightarrow \psi + \ell^{+}\ell^{-}$ phase space is to be identified as
photon fragmentation, and what part is interpreted as lepton fragmentation. In
Eq. (\ref{eq:lofrag}), the dependence on the factorization scale $\mu$ cancels
between $D_{\ell \rightarrow \psi}$ and $d \widehat{\Gamma} / dE_{\psi}$. Thus
by changing $\mu$ some of the lepton fragmentation contribution can be moved
into the photon fragmentation term and vice versa. There is therefore no clear
crossover from lepton fragmentation to photon fragmentation, which makes it
necessary to make an arbitrary choice on the appropriate phase space cutoff.
In this paper a cutoff on the invariant mass of the $\ell - \psi$ system,
where $\ell$ is the fragmenting lepton, is introduced. The
contribution to the differential decay rate from negative lepton
fragmentation is considered to come from the region of phase space
where $s<\mu^{2}$. Here $s$ is the invariant mass of the $\ell^{-} -
\psi$ system. Similarly the contribution from positive lepton
fragmentation is considered to come from the region $s'<\mu^{2}$ where
$s'$ is the invariant mass of the $\ell^{+} - \psi$ system. The photon
fragmentation contribution is interpreted as coming from the remaining
region. In the calculation of the photon subprocess the invariant-mass
cutoffs translate into a limit on the phase space of the photon energy
distribution.

The decay rate for $Z^{0}\rightarrow \gamma +\ell^{+}\ell^{-}$ is
\begin{equation}
\Gamma(Z^{0}\rightarrow\gamma + \ell^{+}\ell^{-})=\frac{1}{2M_{Z}}
\int[dk][dp_{+}][dp_{-}] \; (2\pi)^{4}\delta^{4}(Z-k-p_{+}-p_{-})
\; \frac{1}{3}\sum |A|^{2},
\label{eq:pr}
\end{equation}
where $k$, $p_{+}$, $p_{-}$ and $Z$ are the 4-momenta of the photon,
$\ell^{+}$, $\ell^{-}$, and $Z^{0}$. The amplitude can be calculated from the
diagrams in figure 4. Averaging over initial spins and summing over final
spins reduces the square of the amplitude to
\begin{eqnarray}
 \frac{1}{3}\sum |A|^{2} & = & \frac{g_{w}^{2}e^{2}}{6 \cos^{2}\theta_{w}}
\; (C_{V}^{2}+C_{A}^{2}) \; \frac{(y-2)^2+w^{2}}{y^{2}-w^{2}}
\label{eq:pa}
\end{eqnarray}
where $y=2 k \cdot Z/M^{2}_{Z}$, and $w=2(p_{+}-p_{-}) \cdot Z/M^{2}_{Z}$,
$\; C_{V}=-1+4 \sin^{2}\theta_w$ and $C_{A}=1$, $\;g_{w}$ is the weak
coupling constant, and $\theta_{w}$ is the weak mixing angle. Note that aside
{}from the normalization this is the same as the integrand of
Eq. (\ref{eq:rate}) in the limit $M_{\psi} \rightarrow 0$. Simplifying the
phase space integral, Eq. (\ref{eq:pr}) reduces to
\begin{equation}
\Gamma(Z^{0}\rightarrow\gamma + \ell^{+}\ell^{-})=
\frac{M_{Z}}{32 (2\pi)^3}\int^{1}_{2 \mu^{2}/M^{2}_{Z}} dy
\int^{( y-2\mu^{2}/M_{Z}^{2} )}_{-( y-2\mu^{2}/M_{Z}^{2})} dw
\; \frac{1}{3}\sum |A|^{2},
\label{eq:pps}
\end{equation}
where the limits on the invariant masses $s$ and $s'$ translate into
the limits on $w$. Integrating over $w$ gives the photon energy
distribution in the decay $Z^{0}\rightarrow\gamma+\ell^{+}\ell^{-}$:
\begin{eqnarray}
\lefteqn{\frac{d \widehat{\Gamma}}{dE_{\gamma}}(Z^{0}\rightarrow
\gamma(E_{\gamma}) + \ell^{+}\ell^{-},\mu^2)=}
\nonumber \\
& &
\frac{2 \alpha}{\pi} \; \frac{\Gamma(Z^{0}\rightarrow \ell^{+}\ell^{-})}{M_{Z}}
\; \left[ \frac{(y-1)^{2} +1}{y}\log{\left( \frac{y-\mu^{2}/M^{2}_{Z}}
{\mu^{2}/M^{2}_{Z}} \right)} \; - \; y \; + \; 2\frac{\mu^{2}}{M_{Z}^{2}}
\right] \; \theta (y-\frac{2 \mu^{2}}{M^{2}_{Z}}),
\label{eq:dpr}
\end{eqnarray}
where $y = 2 E_{\gamma}/M_{Z}$. It is possible to simplify the expression for
the energy distribution by taking the limit $\mu \ll M_{Z}$. In this limit
Eq. (\ref{eq:dpr}) simplifies to
\begin{equation}
\frac{d \widehat{\Gamma}}{dE_{\gamma}}(Z^{0}\rightarrow
\gamma(E_{\gamma}) + \ell^{+}\ell^{-},\mu^2) =
\frac{2 \alpha}{\pi} \; \frac{\Gamma(Z^{0}\rightarrow \ell^{+}\ell^{-})}{M_{Z}}
\; \left[ \frac{(y-1)^{2} +1}{y}\log{\frac{yM^{2}_{Z}}{\mu^{2}}} \;
- \; y \right].
\label{eq:dpr2}
\end{equation}
The price that is paid for this simplification is that smooth
threshold behavior at $y =2 \mu^{2}/M^{2}_{Z}$ is lost, and the differential
decay rate becomes negative at sufficiently small values of $y$. Since the
fragmentation approximation breaks down in the threshold region anyway
nothing is lost by making this simplification.

\bigskip
{\bf\centerline{Lepton Fragmentation}}

The calculation of the fragmentation function
$D_{\ell\rightarrow\psi}(z,\mu^{2})$ for a lepton to split into a $\psi$
parallels that of the photon fragmentation function. At leading order in
$\alpha$ the process is symbolically represented by the diagram in figure 5a.
The fragmentation probability is obtained by
dividing $\sigma_{1}$, the cross section for the production of
$\psi + \ell^{-}$, by $\sigma_{0}$, the cross section for lepton production
shown in the diagram of figure 5b, in the limit $E_{\psi} \gg M_{\psi}$ where
fragmentation dominates.

The general form of the cross section $\sigma_{0}$ for lepton production is
\begin{equation}
\sigma_{0} = \frac{1}{Flux} \; \int [dq][dp_{out}] \; (2 \pi)^{4}
\delta^{4} (p_{in}-q-p_{out}) \; \sum|A_{0}|^{2},
\label{eq:lgz}
\end{equation}
where $q$, $p_{in}$, and $p_{out}$ are the 4-momenta of the $\ell^{-}
$, the incoming particles, and all other outgoing particles. Just as
in the photon fragmentation calculation, $[dq]$ and $[dp_{out}]$ are
the Lorentz invariant phase space for the lepton and the remaining
outgoing particles, and $Flux$ represents the incoming particle flux
(which will cancel with the same quantity in the cross section
$\sigma_{1}$). The square of the amplitude $A_{0}$ calculated from the
Feynman diagram in figure 4b for $\ell^{-} $ production, averaged over
initial spins and summed over final spins, is
\begin{equation}
\sum |A_{0}|^{2}={\rm tr}[ \not{\! q} \; \Gamma \bar{\Gamma}]
\label{eq:la}
\end{equation}
where the Dirac matrix $\Gamma$ is the matrix element for the production of a
real lepton of momentum $q$ for which the explicit form is not needed. The
lepton mass $m_{\ell}$ has been neglected since its 4-momentum $q$ is
taken to be large compared to $m_{\ell}$.

The general form of the cross section $\sigma_{1}$ for the production of a
lepton that subsequently fragments into a $\psi$ is
\begin{equation}
\frac{1}{Flux} \; \int [dP][dp_{-}][dp_{out}] \; (2 \pi)^{4}
\; \delta^{4} (p_{in}-P-p_{-}-p_{out}) \;
\sum|A_{1}|^{2},
\label{eq:l}
\end{equation}
where $P$ is the $\psi$ 4-momentum, and everything else is as
described after Eq. (\ref{eq:lgz}).  The next step is to write the
phase space in an iterated form, by introducing integrals over
$q=P+p_{-}$ the virtual lepton momentum, and $s=q^{2}$ the invariant
mass of the $\psi - \ell^{-}$ system. Then the phase space expression
becomes
\begin{eqnarray}
&& \int [dP] [dp_{-}] \; (2 \pi)^4 \delta^4(p_{in} - P - p_{-} - p_{out})
\nonumber \\
&=& \int {d s \over 2 \pi} \int [dq]
        \; (2 \pi)^4 \delta^4(p_{in} - q - p_{out})
        \int [dP] [dp_{-}] \; (2 \pi)^4 \delta^4(q - P - p_{-}) \;.
\label{eq:itps}
\end{eqnarray}
The contribution that corresponds to the fragmentation of the lepton in the
diagram of figure 5a comes from the region of phase space in which the
$\psi-\ell^{-}$ system has large momentum $q$ compared to the $\psi$
mass and small invariant mass $s=q^{2}$ of order $M_{\psi}^{2}$. In a frame in
which the virtual lepton has a 4-momentum $q=(q_{0},0,0,q_{3})$, the
longitudinal momentum fraction of the $\psi$ relative to the $\psi-\ell^{-}$
system is $z=(P_{0}+P_{3})/(q_{0}+q_{3})$ and its transverse momentum is
$\vec{P}_{\perp}=(P_{1},P_{2})$. Expressing the phase
space in terms of these variables and integrating over the 4-momentum $p_{-}$
and over $\vec{P}_{\perp}$, the 2-body phase space reduces to \cite{bcy:cfrag}
\begin{equation}
\int[dP][dp_{-}] \; (2\pi)^{4}\delta^{4}(q-P-p_{-})=\frac{1}{8\pi}\int^{1}_{0}
\! dz \; \theta \left( s-\frac{M_{\psi}^2}{z} \right).
\label{eq:tbps}
\end{equation}
The lepton mass has been set to zero. An  upper limit on the integral
over the invariant mass $s$ is introduced by requiring  $s<\mu^{2}$,
as discussed earlier.

The calculation of the lepton framentation function is simplest to do in the
axial gauge, because only the diagram of figure 5a, where the lepton is
produced and splits into a collinear lepton and $\psi$, needs to be
considered. Other diagrams where both the lepton and $\psi$ are produced
separately from the vertex $\Gamma$ are suppressed in this gauge. If the
calculation were done in some other gauge these diagrams would need to be
considered, but the resulting expression could be manipulated into the form
below using Ward identities. The 4-vector $N$ associated with axial gauge is
chosen to be $N=(1,0,0,-1)$. The amplitude $A_{1}$ calculated in this gauge
can be reduced to
\begin{equation}
A_{1}=e^2 g_{\psi} \;
\epsilon_{\beta}(P)^{*} \; \frac{1}{s} \; (\bar{u}(p_{-}) \, \gamma_{\alpha}
{\not \! q} \, \Gamma) \left (g^{\alpha\beta}-\frac{P^{\alpha}N^{\beta}+
P^{\beta}N^{\alpha}}{P\cdot N} \right).
\label{eq:a1}
\end{equation}
where $\epsilon$ is the $\psi$ polarization vector, and $\Gamma$ is the
Dirac matrix element for the production of a virtual lepton with an invariant
mass $s$ on the order of the $\psi$ mass. The explicit form of $\Gamma$ is not
needed.

The $\psi$ 4-momentum can be written as $\; P^{\mu}=zq^{\mu}+P^{\mu}_{\perp}+
(P_{0}q_{3}-q_{0}P_{3})/(q_{0}+q_{3}) N^{\mu}$. In the fragmentation region
$P_{\perp}^{\mu}=(0,\vec{P}_{\perp}, 0)$, and $P_{0}q_{3}-q_{0}P_{3}$ are of
order $M_{\psi}$ while the virtual lepton momentum $q$ is large compared to
$M_{\psi}$ so $P \simeq zq$. Note that in the fragmentation region both
$s=q^{2}$ and $P\cdot q$ are of order $M^{2}_{\psi}$. Using these
approximations and keeping only the leading order terms in
${\not \! q}/M_{\psi}$ simplifies the square of the amplitude to
\begin{equation}
\sum|A_{1}|^{2}= 32 \pi^{2} \alpha^{2} g^{2}_{\psi}
\left(\frac{(z-1)^{2}+1}{z} \; \frac{1}{s} \; - \; \frac{M_{\psi}^{2}}
{s^{2}}\right){\rm tr}[{\not \! q} \; \Gamma \bar{\Gamma}].
\label{eq:ssa1}
\end{equation}
Integrating over $s$ up to the scale $\mu^{2}$ the
lepton fragmentation probability is obtained by dividing the cross section
$\sigma_{1}$ by the cross section $\sigma_{0}$. The differences between
$\sigma_{1}$ and $\sigma_{0}$ are due to the fact that
$q^{2} \sim M^{2}_{\psi}$ in $\sigma_{1}$, while $q^{2}=0$ in $\sigma_{0}$.
These differences are on the order of $M^{2}_{\psi}/E^{2}_{\psi}$ and in the
fragmentation limit where $E_{\psi} \gg M_{\psi}$ they can be neglected.
The result is
\begin{equation}
\frac{\sigma_{1}}{\sigma_{0}} = \int^{\mu^{2}}_{0} ds \; \int^{1}_{0} dz
\; \theta(s-\frac{M^{2}_{\psi}}{z}) \; 2 \alpha^{2} g^{2}_{\psi} \left(
\frac{(z-1)^{2}+1}{z} \;\frac{1}{s} \; - \; \frac{M^{2}_{\psi}}{s^{2}} \right).
\label{eq:int1}
\end{equation}
{}from this it is possible to extract the lepton fragmentation function
\begin{equation}
D_{\ell\rightarrow\psi}(z,\mu^{2}) =  2 \alpha^{2} g^{2}_{\psi}
\left[ \frac{(z-1)^{2}+1}{z} \log{\frac{z \mu^{2}}{M_{\psi}^{2}}}
\; - \; z \; + \; \frac{M_{\psi}^{2}}{\mu^{2}} \right] \; \theta (z
\mu^{2} - M^{2}_{\psi})
\label{eq:lff}
\end{equation}
Note that the lepton fragmentation function is zero for values of $z$
at which the production of a $\psi$ is kinematically forbidden.
Taking the limit $\mu \gg M_{\psi}$ simplifies Eq. (\ref{eq:lff}) to
\begin{equation}
D_{\ell\rightarrow\psi}(z,\mu^{2})  \approx  2 \alpha^{2} g^{2}_{\psi}
\left[ \frac{(z-1)^{2}+1}{z} \log{\frac{z \mu^{2}}{M_{\psi}^{2}}}
\; - \; z\right].
\label{eq:lff2}
\end{equation}
Just as in the calculation of the photon subprocess there is a price to be
paid for this simplification. Eq. (\ref{eq:lff2}) does not have the correct
threshold behavior at $z=M^{2}_{\psi}/\mu^{2}$, and it becomes negative for
sufficiently small $z$. The fragmentation function Eq. (\ref{eq:lff2}) is
shown at the scales $\mu=3M_{\psi}$ and $\mu=6M_{\psi}$ in figure 6. Note that
there is a dramatic dependence on the arbitrary factorization scale $\mu$. At
the scale $\mu = 6M_{\psi}$ the fragmentation function is much larger and peaks
at a lower $z$ value than at the scale $\mu=3M_{\psi}$.

\bigskip
{\bf\centerline{Comparison with full calculation}}

The fragmentation contribution to the differential decay rate for
$Z^{0}\rightarrow\psi +\ell^{+}\ell^{-}$ is given by inserting
$P_{\gamma \rightarrow \psi}$ from Eq. (\ref{photprob}), $d
\widehat{\Gamma}/d E_{\psi}$ from Eq. (\ref{eq:dpr2}), and $D_{\ell
\rightarrow \psi}$
{}from Eq. (\ref{eq:lff2}) into the factorization
formula Eq. (\ref{eq:lofrag}):
\begin{equation}
\frac{d \Gamma}{d E_{\psi}}(Z^{0} \rightarrow \psi(E_{\psi})+
\ell^{+}\ell^{-}) =
8 \alpha^{2} g^{2}_{\psi}
\frac{\widehat{\Gamma}(Z^{0}\rightarrow \ell^{+}\ell^{-})} {M_{Z}}
\left[ \frac{(y-1)^{2}+1}{y}\log{\frac{y^{2} M^{2}_{Z}}{M^{2}_{\psi}}} \;
- \; 2y \right].
\label{fragcon}
\end{equation}
Note that the $\mu$-dependence cancels exactly.
It is now possible to verify that this agrees with the full calculation
in the fragmentation limit Eq. (\ref{xxx}).

Figure 7 compares the energy distribution of the full calculation Eq.
(\ref{eq:gfun}) and the fragmentation calculation Eq.
(\ref{fragcon}). It is clear from the graph that the fragmentation
approximation breaks down for sufficiently small $E_{\psi}$. For
$E_{\psi}=3M_{\psi}$ the difference between the two curves is less than
$1 \%$, while for $E_{\psi}=2M_{\psi}$ the difference is  $5 \%$. In practice
there will often be a minimum energy below which detectors do not register
particles. If this minimum energy is large enough then it is clear that the
fragmentation approximation will give a result very close to the full
calculation.

Figure 8 shows the energy distribution in the fragmentation limit
separated into the lepton fragmentation contribution, the first term
on the right hand side of Eq. (\ref{eq:lofrag}), and the photon
fragmentation contribution, the last term on the right hand side of
Eq. (\ref{eq:lofrag}). The contributions are shown at $\mu=3M_{\psi}$
and $\mu=6M_{\psi}$. The relative contribution of the two production mechanisms
depends dramatically on the factorization scale $\mu$, though the total, photon
fragmentation plus lepton fragmentation is independent of $\mu$. At the scale
$\mu = 3 M_{\psi}$ the contribution of the lepton fragmentation mechanism is
negligable compared to the contribution from the photon fragmentation
mechanism, while at the scale $\mu = 6 M_{\psi}$ both contributions are of the
same order.

The analysis carried out in this paper applies equally to other heavy quark
anti-quark states such as the $\psi '$ and the $\Upsilon$. Unfortunately the
framentation contribution to $\Upsilon$ production is an order of magnitude
smaller than the fragmentation contribution to $\psi$ production. This is
because the $\Upsilon$-photon coupling $g^{2}_{\Upsilon}=6 \times 10^{-4}$
is so much smaller than the $\psi$-photon coupling
$g^{2}_{\psi}=8 \times 10^{-3}$, and the $\Upsilon$ mass is larger than the
$\psi$ mass.

\bigskip
{\bf\centerline{Conclusion}}

The process $Z^{0} \to \psi + \ell^{+} \ell^{-}$ has been studied in the
fragmentation limit $M_{Z} \to \infty$ keeping $E_{\psi}/M_{Z}$ fixed. In this
limit the decay rate factors into the subprocess rates
$\widehat{\Gamma}(Z^{0} \to \ell^{+}\ell^{-})$ and
$\widehat{\Gamma}(Z^{0} \to \gamma + \ell^{+} \ell^{-})$, convoluted with
the electromagnetic fragmentation functions
$D_{\ell \to \psi}$ and $D_{\gamma \to \psi}$. The fragmentation function
$D_{\ell \rightarrow \psi}(z, \mu)$ for a lepton to split into $\psi$, the
fragmentation probability $P_{\gamma \rightarrow \psi}$ for a photon to split
into $\psi$, and the subprocess
$\widehat{\Gamma}(Z^{0} \to \gamma + \ell^{+} \ell^{-})$ were calculated at
lowest order in $\alpha$. The lepton
fragmentation function was defined by imposing a cutoff on the invariant mass
$s$ of the lepton and $\psi$ in the final state. This cutoff translated into
limits on the phase space of the subprocess
$\widehat{\Gamma}(Z^{0} \to \gamma + \ell^{+} \ell^{-})$. It was then
explicitly shown that the $\mu$-dependence of
the lepton fragmentation function canceled the $\mu$-dependence of the
subprocess $\widehat{\Gamma}(Z^{0} \to \gamma + \ell^{+} \ell^{-})$. Comparison
between the fragmentation calculation and the full calculation shows that the
fragmentation approximation is accurate to within $5 \%$ at $E_{\psi} =
2M_{\psi}$, and becomes more accurate for $\psi$ energies greater than this.

This work is supported in part by the U.S. Department of Energy, Division of
High Energy Physics, under Grant DE-FG02-91-ER40684. I wish to thank E. Braaten
for many helpful discussions, and his infinite patience. I also wish to thank
the Fermilab theory group for their hospitality.

\vfill\eject

\noindent{\Large\bf Figure Captions}
\begin{enumerate}
\item The Feynman diagrams for $Z^{0} \rightarrow \psi + \ell^{+} \ell^{-}$
        at leading order in $\alpha$. a) The two diagrams from which the
        fragmentation contribution can be isolated, and  b) the two diagrams
        that may be neglected.
\item The two Feynman diagrams for $Z^{0} \rightarrow \psi \gamma$ at leading
        order in $\alpha$.
\item Feynaman diagrams for a) $\psi$ production by photon fragmentation,
         b) photon production. The
         shaded circle represents some general vertex that radiates the photon.
\item The two Feynman diagrams for $Z^{0} \rightarrow \gamma + \ell^{+}
        \ell^{-}$ at leading order in $\alpha$.
\item Feynman diagrams for a) $\psi$ production by lepton fragmentation,
         b) lepton production. The
         shaded circle represents some general vertex that radiates the
         lepton.
\item Lepton fragmentation function for $\mu=3M_{\psi}$ (solid) and
        $\mu=6M_{\psi}$ (dashes).
\item Energy distribution: the full calculation (solid), the fragmentation
        calculation (dashes).
\item Photon fragmentation (PF) and lepton fragmentation (LF) contributions to
        the energy distribution for $\mu=3M_{\psi}$ (solid) and
        $\mu=6M_{\psi}$ (dashes).
\end{enumerate}

\vfill\eject

\end{document}